\documentclass[a4paper,11pt]{article}
\pdfoutput=1 % if your are submitting a pdflatex (i.e. if you have
\usepackage{jheppub} % for details on the use of the package, please
\usepackage[T1]{fontenc} % if needed

\title{\boldmath Error correcting codes and heterotic Narain CFTs}

\author[a,b%,1
]{Shun'ya Mizoguchi\note{Corresponding author.}
}
\author[b]{and Takumi Oikawa}
\affiliation[a]{Theory Center, Institute of Particle and Nuclear Studies,
KEK, \\Tsukuba, Ibaraki, 305-0801, Japan }
\affiliation[b]{SOKENDAI (The Graduate University for Advanced Studies)\\
Tsukuba, Ibaraki, 305-0801, Japan }
\emailAdd{mizoguch@post.kek.jp}
\font\mybb=msbm10 at 12pt

\font\mybbsub=msbm10 at 8pt
\font\mybbsmall=msbm10 at 10pt

\font\myeu=eufm10 at 12pt

\font\myeusub=eufm10 at 8pt
\font\myeusmall=eufm10 at 10pt

\def\bb#1{\hbox{\mybb#1}}

\def\bbsub#1{\hbox{\mybbsub#1}}
\def\bbsmall#1{\hbox{\mybbsmall#1}}

\def\frak#1{\hbox{\myeu#1}}

\def\fraksub#1{\hbox{\myeusub#1}}
\def\fraksmall#1{\hbox{\myeusmall#1}}

\def\ZZ {\bb{Z}}
\def\ZZsub {\bbsub{Z}}

\def\CCsub {\bbsub{C}}

\def\FF {\bb{F}}
\def\FFsub {\bbsub{F}}
\def\FFsmall {\bbsmall{F}}
\def\g {\frak{g}}
\def\gsub {\fraksub{g}}
\def\gsmall {\fraksmall{g}}

\newcommand\beqa{\begin{eqnarray}}
\newcommand\eeqa{\end{eqnarray}}
\newcommand\n{\nonumber\\}

\abstract{
We study error correcting codes that construct the Narain lattices of 
heterotic strings as code lattices. 
We identify, in both $E_8\times E_8$ 
and Spin$(32)/\ZZ_2$ heterotic strings, a pair of a binary code 
and a set of the corresponding metric, B field, and background gauge field, 
such that the lattice constructed from the binary code by Construction A 
coincides with the Narain lattice.
We also construct heterotic Narain lattices using codes over $\FF_3$ 
and $\FF_5$ by Construction A${}_{\CCsub}$ and 
``Construction A${}_{\gsmall}$'' with $\g=SU(5)$, respectively.
As a bi-product, we also clarify the relationship between 
codes that construct Euclidean even self-dual lattices and NSR-fermions, 
where the $\ZZ_2$ inversion structure of the generator matrices plays 
a significant role.
}

\begin{document}

\maketitle
\flushbottom

\section{Introduction}
In modern quantum information technology, quantum error correcting 
codes are an indispensable technology to ensure the reliability of 
quantum information (see \cite{Terhal} for a review).
Quantum error correction restores correct information from information containing errors 
that occur when transmitting quantum states such as qubits, 
which act as unitary operators \cite{Shor9qubit,CSScode1,CSScode2,Gottesmanstabilizercode}. 
On the other hand, it has been known for a long time that errors that occur 
when simply transmitting a bit string consisting of $\{0,1\}$ can be removed 
by adding redundancy to the original information bit string \cite{MacWilliamsSloane}. 
This is called classical error correction.

As is well known, classical error correcting codes are closely related to 
the sphere packing problem \cite{ConwaySloan}.
In error correcting codes, the information to be sent is represented 
as one of a set of spatially arranged points.
The essence of error correcting codes is that even if the position of 
a point shifts slightly from its intended location due to some disturbance, 
it can be returned to its original position 
by looking for the point closest to that point where the information 
should originally be located.
Therefore, placing the points as far away from each other as possible 
will lead to higher correction ability.
On the other hand, however, separating the points means providing 
wasted space for information other than the original information, 
which is not preferable from the viewpoint of efficiency of information transmission.
Therefore, the problem is how to arrange the information 
as small as possible while keeping the distance between the pieces of information.
This is the reason why classical error correcting codes 
and the sphere packing problem are related.

It is known \cite{ConwaySloan} that the eight-dimensional close-packed lattice is the $E_8$ lattice, 
and the 24-dimensional close-packed lattice is the Leech lattice  
as proven recently in Refs.\cite{Viazovska,Cohnetal}.
%[https://arxiv.org/abs/1611.01685].
These lattices have long been familiar in the study of string theory 
\cite{heterotic_string1,heterotic_string2,heterotic_string3,Narain1986,Narain1987}
and vertex operator algebra \cite{Borcherds,FLM}, and the theta functions associated with them exhibit good modular transformation properties.
The relationship between classical error correcting codes and the modular 
transformation properties of theta functions associated with lattices created by the so-called 
Construction A is also well known \cite{MacWilliams,ConwaySloan,Ebeling}.

Recently, it has been discovered that certain {\em quantum} error-correcting codes can also 
be mapped onto classical error-correcting codes \cite{Dymarsky2020}, 
and that the momentum-winding lattice of the Narain CFT can be realized as 
a lattice with an indefinite metric constructed from it (see \cite{Dymarsky2020,
Dymarsky:2020bps,Dymarsky:2020pzc,Dymarsky2021,
Buican:2021uyp,Yahagi2022,Furuta:2022ykh,Henriksson2023,
Angelinos:2022umf,Henriksson:2022dml,Dymarsky:2022kwb,
Kawabata2023a,Furuta2023,Alam2023,Kawabata2023b,
Kawabata2024a,Buican:2023bzl,Aharony:2023zit,Barbar:2023ncl,Buican:2023ehi,Kawabata2024b,
MizoguchiOikawa,Kawabata:2025hfd,Angelinos:2025mjj,
Ando:2025hwb,Dymarsky:2025agh,Keller:2025elq}
for an incomplete list on this subject).
However, Narain lattices constructed so far in this way are limited to those with signature $(d,d)$.
Since the Narain CFT was originally considered/discovered in the compactification 
of heterotic strings with a lattice signature $(16+d,d)$, the following question naturally arises:
What kind of error-correcting code constitutes this original Narain lattice?
The purpose of this paper is to answer this question.

We first realize the Narain lattice for heterotic strings as 
a code lattice of a binary code.
Using already known results on the code that constructs the Narain lattice 
of bosonic strings and the code that constructs the 16-dimensional even self-dual 
gauge lattice of the heterotic string, 
we determine the generator matrix of the code 
and the corresponding heterotic background 
such that the code lattice coincides with the heterotic Narain lattice. 
This is done by carefully comparing the generator matrices of the codes with 
those of the momentum-winding lattice of heterotic strings.
Next, we utilize the fact that the $E_8$ lattice can be constructed from 
the tetracode by Construction A${}_{\CCsub}$, 
and combine this with our previous paper's construction \cite{MizoguchiOikawa}
of a Narain lattice for bosonic strings 
using codes over $\FF_3$ to realize a heterotic Narain lattice as a code lattice of 
a code over $\FF_3$.
Furthermore, using the fact that the $E_8$ lattice can also be realized 
as a lattice of integers of the cyclotomic field with $m=5$, 
we apply the results of \cite{MizoguchiOikawa} to realize a  
heterotic Narain lattice as a code lattice of a code over $\FF_5$.
In \cite{MizoguchiOikawa}, this lattice construction is rephrased 
in terms of Lie algebras and called Construction A${}_{\gsmall}$ (here $\g=SU(5)$).
As will be explained in more detail in the main text,
this is essentially  gluing together root lattices using a code 
over the quotient ring of the weight lattice by the root lattice of the Lie algebra.

This paper is organized as follows. Section 2 briefly summarizes the Narain lattice 
obtained by compactifying heterotic strings. 
In Section 3, we set the metric, B field, 
and gauge field on the compact space
and show that a lattice constructed by Construction A from a code determined by 
them is indeed a Narain lattice of a heterotic string. 
Section 4 considers the compactifications on a one- and two-dimensional tori 
as concrete examples of the results obtained in Section 3.
In Sections 5 and 6, we construct heterotic Narain lattices using codes over $\FF_3$ and $\FF_5$ 
by Construction A${}_{\CCsub}$ and Construction A${}_{\gsmall}$ with $\g=SU(5)$, respectively.
Finally, Section 7 concludes and summarizes the discussion.
In Appendix, we summarize the relationship between the codes 
that construct the $D_{16}^+$ lattice and the NSR fermions.

\section{Heterotic Narain lattice}
The general partition function of a two-dimensional scalar-conformal field 
theory compactified on a $d$-dimensional torus is given by
\cite{Narain1986,Narain1987}
\beqa
{\rm Tr}\,q^{L_0-\frac c{24}}\bar q^{\bar L_0-\frac c{24}}
&=&
\frac1{|\eta(\tau)|^{2c}}
\sum_{n_i\in\ZZsub^d}\sum_{w^i\in\ZZsub^d}
q^{\frac12 p_L^\alpha p_{L\alpha}}
\bar q^{\frac12 p_R^\alpha p_{R\alpha}},\\
p_{L\,\alpha}&=&
\sqrt{\frac{\alpha'}2}
\left(
 \frac{n_i}R+\frac2{\alpha'}\frac{B_{ij}+g_{ij}}2 w^j R
\right)e^i_{~\alpha},\\
p_{R\,\alpha}&=&
\sqrt{\frac{\alpha'}2}
\left(
 \frac{n_i}R+\frac2{\alpha'}\frac{B_{ij}-g_{ij}}2 w^j R
\right)e^i_{~\alpha}
\eeqa
where the central charge $c$ is $d$.
This can be seen as the partition function of bosonic conformal field theory 
whose target space is a compact $d$-dimensional torus in the ten-dimensional 
spacetime of superstring.
On the other hand, in the heterotic string, 16-dimensional left moving bosons 
$X^I$ $(I=1,\ldots,16)$ also contribute to the partition function.
$X^I$ is coupled to the background gauge field $A_\mu^I$ of rank $16$, 
and the partition function in this case is given by \cite{Ginsparg}
\beqa
{\rm Tr}\,q^{L_0-\frac c{24}}\bar q^{\bar L_0-\frac c{24}}
&=&
\frac1{|\eta(\tau)|^{2c}}
\sum_{n_i\in\ZZsub^d}\sum_{w^i\in\ZZsub^d}
\sum_{V^I\in\Lambda_{\rm int}}
q^{\frac12 p_L^\alpha p_{L\alpha}
+\frac12 p_L^I p_L^I}\,
\bar q^{\frac12 p_R^\alpha p_{R\alpha}},
\\
p_{L\,\alpha}&=&
\sqrt{\frac{\alpha'}2}
\left(
\frac{n_i}R
-V^I A_i^I
+
\left(\frac2{\alpha'}\frac{B_{ij}+g_{ij}}2 
-\frac{A_i^I A_j^I}2 \right)w^j R
\right)e^i_{~\alpha},
\\
p_{R\,\alpha}&=&
\sqrt{\frac{\alpha'}2}
\left(
\frac{n_i}R
-V^I A_i^I
+
\left(\frac2{\alpha'}\frac{B_{ij}-g_{ij}}2 
-\frac{A_i^I A_j^I}2 \right)w^j R
\right)e^i_{~\alpha},\\
p^I&=&
%\sqrt{\frac{\alpha'}2}
%(
V^I+w^i RA_i^I
%)
\label{p_L^I}
\eeqa
where $V^I\in\Lambda_{\rm int}$ $(I=1,\ldots,16)$ 
is a vector representing a lattice point on a 
$16$-dimensional internal Euclidean self-dual lattice.
We set $\alpha'=2$ hereafter.

Following the convention of the construction of the Narain lattice 
by Construction A, we define an orthogonal transformation of the 
momentum-winding lattice  $(p_{L\,\alpha},p_{R\,\alpha},p^I)$ as 
\beqa
(\lambda_{1\,\alpha},\lambda_{2\,\alpha},p^I)=
\left(\frac{p_{L\,\alpha}-p_{R\,\alpha}}{\sqrt 2},
\frac{p_{L\,\alpha}+p_{R\,\alpha}}{\sqrt 2},
p^I\right).
\eeqa
This lattice is even self-dual with respect to the inner product defined by 
the matrix \footnote{Of course, this $\eta$ 
should not be confused with the Dedekind $\eta$ 
that appears in the partition function.}
\beqa
\eta&=&
\left(
\begin{array}{ccc}
  0& I_d  & 0  \\
  I_d& 0  & 0  \\
0  &  0 & I_{16}  
\end{array}
\right),
\label{innerproducteta}
\eeqa
where $I_m$ denotes the $m\times m$ identity matrix.
Specifically, $\lambda_{1\,\alpha}$ and $\lambda_{2\,\alpha}$ are given as follows:
\beqa
\lambda_{1\,\alpha}&=&
\frac R{\sqrt{2}}g_{ij}w^j e^i_{~\alpha},\n
\lambda_{2\,\alpha}&=&
\left(\frac {\sqrt{2}}R n_i 
-\sqrt{2}V^I A^I_i
+\frac R{\sqrt{2}}(B_{ij}-A_i^IA_j^I)w^j
\right)
e^i_{~\alpha},
\eeqa
where $e^i_{~\alpha}$ is the inverse vielbein satisfying
\beqa
e^i_{~\alpha}e^j_{~\beta}\delta^{\alpha\beta}&=&g^{ij}.
\eeqa
$(\lambda_{1\,\alpha},\lambda_{2\,\alpha},p^I)$ are invariant under the scaling:
\beqa
&&R\rightarrow \lambda R,\n
&&g_{ij}\rightarrow \lambda^{-2} g_{ij},~~~e^i_{~\alpha}\rightarrow \lambda e^i_{~\alpha},  \n
&&B_{ij}\rightarrow \lambda^{-2} B_{ij},\n
&&A_{i}^{^I}\rightarrow \lambda^{-1} A_{i}^{^I}.
\label{scaling}
\eeqa

In matrix notation they become
%\beqa
%(\lambda_{1\,\alpha},\lambda_{2\,\alpha},p^I)
%&=&
%(w^i,n_i,v^A)
%%
%\left(
%\begin{array}{ccc}
% \frac R{\sqrt{2}}\underbrace{G_{ji}e^j_{~\alpha}}_{=e_{i\alpha}} & \frac R{\sqrt{2}}(B_{ji}-A_j^IA_i^I)e^j_{~\alpha}  &  RA_i^{~I} \\
% 0 & \frac {\sqrt{2}}R  e^i_{~\alpha} &  0 \\
% 0 & -\sqrt{2}(\gamma_{16})_A^{~I} A^I_{~j}e^j_{~\alpha} &   (\gamma_{16})_A^{~I}
%\end{array}
%\right)
%%
%\eeqa
\beqa
(\vec \lambda_1,\vec \lambda_2,\vec p)
&=&
(\vec w,\vec n,\vec v)
\left(
\begin{array}{ccc}
 \frac R{\sqrt{2}}\gamma & \frac R{\sqrt{2}}(B^T-A\,A^T)\gamma^{T-1}   &  RA \\
 0 & \frac {\sqrt{2}}R  \gamma^{T-1} &  0 \\
 0 & -\sqrt{2}\gamma_{16} A^T\gamma^{T-1}  &   \gamma_{16}
\end{array}
\right).
\label{matrix_representation}
\eeqa
Here we have used the basis $\alpha_A^{~I}$ $(A=1,\ldots,16)$ 
of $\Lambda_{\rm int}$ to expand $V^I$ as
\beqa
V^I&=&v^A\alpha_A^{~I}~~~(v^A\in\ZZ^{16}),
\eeqa
and defined $16\times 16$ matrix $\gamma_{16}$ as
\beqa
(\gamma_{16})_A^{~~I}:=\alpha_A^{~~I}.
\eeqa
$\gamma$ and $ \gamma^{T-1}$ are $d\times d$ matrices 
\beqa
(\gamma)_i^{~\alpha}:=e_i^{~\alpha},~~~(\gamma^{T-1})^i_{~\alpha}:=e^i_{~\alpha}.
\eeqa
$B$ is also a $d\times d$ matrix whose entries are 
the components of the $B$ field
\beqa
(B)_{ij}:=B_{ij},
\eeqa
whereas $A$ is a $d\times 16$ matrix whose entries are 
the components of the gauge field
\beqa
(A)_i^{~I}:=A_i^{~I}.
\eeqa
$\vec n,\vec w\in\ZZ^{d}$ are the quantized momentum and winding number.

\section{Heterotic Narain lattice constructed from binary codes
}
Let us examine what code over $\FF_2$ constructs this Narain lattice.
In the case of bosonic string, 
when compactified to a square torus of radius $R=1$, 
the Narain lattice corresponds directly to a lattice constructed from 
a ``$B$-form code'' (= a code in the systematic form) by Construction A.
Therefore, in (\ref{matrix_representation}), 
we consider the generator matrix of the momentum lattice $\Lambda^{\rm Narain}$ 
with setting $R=1$ and $\gamma=\gamma^{T-1}=I_d$:
\beqa
G(\Lambda^{\rm Narain})&:=&
\left(
\begin{array}{ccc}
 \frac 1{\sqrt{2}}I_d & \frac 1{\sqrt{2}}(B^T-A\,A^T)   &  A \\
 0 & \sqrt{2} I_d &  0 \\
 0 & -\sqrt{2}\gamma_{16} A^T &   \gamma_{16}
\end{array}
\right).
\label{GLambda}
\eeqa
We look for a code such that the lattice constructed by construction A 
is identical to the lattice constructed by this $G(\Lambda^{\rm Narain})$.
For this, let us consider a generator matrix 
\beqa
G_{\cal C}&:=&
\left(
\begin{array}{ccc}
I_d &B^T-\frac12 A'{A'}^T  &  A' \\
 0 & 0 &   G_{16}
\end{array}
\right),~~~A':=\sqrt{2}A
\label{GC}
\eeqa
 of a $[2d+16,d+8]$ code over $\FF_2$\footnote{\label{footnote_lattice}In this paper, 
 we denote the generator matrix of the lattice $\Lambda$,
 that is,  a matrix such that, as in (\ref{matrix_representation}), 
 when multiplied by an integer row vector 
 from the left, we obtain a lattice point vector,
 by $G(\Lambda)$. 
 On the other hand, the generator matrix of the code ${\cal C}$ is denoted as 
 $G_{\cal C}$ to distinguish it from the generator matrix of the lattice.
%?????}??????_??????????%???????i??q$\Lambda$????????????????????s?????}??A????????????????%
%(\ref{matrix_representation})??????????????????????????????s?????}????????????????????????????????s??x??N??g???????????
%??????????????i??q??_??x??N??g?????????????????????????????????????A???????$G(\Lambda)$???%???\??????B
%??????????A????????${\cal C}$????????????????????s?????}????$G_{\cal C}$????????????????????%
%??\??????A??i??q????????????????????s?????}?????????????????????????}??????????????????B
}.
$G_{\cal C}$ is a $(d+8)\times(2d+16)$ matrix, in which 
$G_{16}$ is the generator matrix of the $[16,8]$ doubly-even self-dual code 
corresponding to $\Lambda_{\rm int}$.
$A$ is chosen so that $A'$ is a $\FF_2$-valued $d\times 16$ matrix.
It should be noted here that the elements of this $G_{\cal C}$ 
take values in $\FF_2$, and $B^T$, $A'$, etc. are abused with 
the same symbols as the background fields of  
heterotic string \footnote{Therefore, since $-1\equiv 1$ mod $2$,
 we can set $B^T=-B\stackrel{!}{=} B$ in $G_{\cal C}$.}.

In the case of binary codes, the code must be doubly-even self-dual 
in order for the lattice constructed by Construction A to be even self-dual,
so the generator matrix must satisfy
 \beqa
 G_{\cal C}\eta G_{\cal C}\equiv 0~~~\mbox{mod $2$}~~~\mbox{
 and}~~~\mbox{diag}G_{\cal C}\eta G_{\cal C}\equiv 0~~~\mbox{mod $4$},
 \eeqa
 where $\eta$ is the inner product matrix (\ref{innerproducteta}).
Since $B^T=-B$, we have
\beqa
 G_{\cal C}\eta G_{\cal C}^T&=&
 \left(
\begin{array}{cc}
0  &   A' {G_{16}}^T \\
 G_{16} {A'}^T &   G_{16}
\end{array}
\right)
\eeqa
so
\beqa
A' {G_{16}}^T\equiv 0~~~\mbox{mod $2$}.
\label{A'G16^T=0mod2}
\eeqa
In addition, for $G_{\cal C}$ (\ref{GC}) to be $\FF_2$-valued, it must also be
\beqa
A'{A'}^T\equiv 0~~~\mbox{mod $2$}.
\label{A'A'^T=0mod2}
\eeqa

Now, the fact that $\Lambda_{\rm int}$ can be constructed from $G_{16}$ using Construction A means that the generator matrix $\gamma_{16}:=G(\Lambda_{\rm int})$ of $\Lambda_{\rm int}$ can be written as 
\beqa
\gamma_{16}&=&\left(
\begin{array}{c}
\frac1{\sqrt 2}G_{16}\\
%S
%
\begin{array}{cc}
0&\sqrt 2 I_8
\end{array}
\end{array}
\right),
%,~~~S:=(O~~\sqrt 2 I_8)
\label{gamma16}
\eeqa
where the $8\times 16$ generator matrix $G_{16}$ of the code is assumed to be 
in systematic form, i.e., the $8\times 8$ part of the left half is 
the identity matrix $I_8$ \footnote{This construction of a lattice generator matrix 
from a code generator matrix is given,
for example, by Conway-Sloan \cite{ConwaySloan}, and(\ref{GLambda}) already uses this idea.
}.
In fact, in this case, the code lattice $\Lambda_{{\cal C}_{G_{16}}}$ constructed from $G_{16}$ by Construction A is
\beqa
\Lambda_{{\cal C}_{G_{16}}}&=&\left\{\left.
\frac{c+2m}{\sqrt 2}\right|c\in{\cal C}_{G_{16}},~m\in\ZZ^{16} 
\right\},
\eeqa
where the part of $\frac{c}{\sqrt 2}$, $c\in{\cal C}_{G_{16}}$ is 
generated by multiplying \\\mbox{$(x_1,x_2,\ldots,x_8)=(x_1,x_2,x_3,x_4,0,0,0,0)$ }
(considered in $\in\FF_2^4$) 
in the lattice generated from $\gamma_{16}$  
\beqa
\Lambda(\gamma_{16})&=&\left\{\left.
(x_1\,x_2,\ldots,x_8)\gamma_{16}\right|(x_1\,x_2,\ldots,x_8)\in\ZZ^8
\right\}
\eeqa
by $\gamma_{16}$ \footnote{
In this paper, given a generator matrix $G$ of a lattice,
we denote the lattice generated by it by $\Lambda(G)$.
Therefore,
in the notation given in footnote \ref{footnote_lattice}, $G=G(\Lambda(G))$.
On the other hand,
when a code generator matrix $G$ is given,
the code $\cal C$ generated by it is denoted by ${\cal C}_G$.
Therefore, again using the notation in footnote \ref{footnote_lattice},
${\cal C}={\cal C}_{G_{\cal C}}$.
Also, given a code ${\cal C}$, we will denote a code lattice constructed from it 
in a specific way, such as Construction A, as $\Lambda_{\cal C}$.
}.
If we consider this in the original $\in\ZZ^4$ instead of $\in\FF_2^4$, 
a shift of $\frac{2m}{\sqrt 2}$ for the eight symbols in the left half of the code is also generated.
Furthermore, the shifts for the eight symbols in the right half can be 
generated by $(x_1\,x_2,\ldots,x_8)S$.
From the above, we found that the code lattice $\Lambda_{{\cal C}_{G_{16}}}$ and
the lattice $\Lambda(\gamma_{16})$ generated from $\gamma_{16}$ are identical.

Therefore, $G(\Lambda^{\rm Narain})$ (\ref{GLambda}) can be written as
\beqa
G(\Lambda^{\rm Narain})&:=&
\left(
\begin{array}{ccc}
 \frac 1{\sqrt{2}}I_d & \frac 1{\sqrt{2}}(B^T-\frac12 A'{A'}^T)   &  \frac 1{\sqrt 2}A' \\
 0 & \sqrt{2} I_d &  0 \\
 0 & -\frac 1{\sqrt 2}G_{16} A'^T &   \frac1{\sqrt 2}G_{16}\\
 0 & -S A'^T &\begin{array}{cc}0&~\sqrt 2 I_8\end{array}
\end{array}
\right),
\label{GLambda2}
\eeqa
and the lattice generated by this $G(\Lambda^{\rm Narain})$ :
\beqa
\Lambda(G(\Lambda^{\rm Narain}))&=&\left\{\left.
(w^1\,\cdots w^d~n_1\,\cdots n_d~v^1\cdots v^{16})G_\Lambda\right|
\right.
\n&&
\left.
~~
(w^1,\,\ldots,w^d;n_1\,\ldots,n_d;v^1,\ldots,v^{16})
\in\ZZ^d\times\ZZ^d\times\ZZ^{16}
\right\}
\eeqa
coincides with the code lattice :
\beqa
\Lambda_{{\cal C}_{G_{\cal C}}}
(=\Lambda_{\cal C})&:=&
\left\{\left.
\frac{c+2m}{\sqrt 2}\right|c\in{\cal C}_{G_{\cal C}}(={\cal C}),~m\in\ZZ^d\times\ZZ^d\times\ZZ^{16}
\right\}
\eeqa
constructed from the code generated by $G_{\cal C}$ (\ref{GC})
by Construction A.

In fact, the first and third rows of $G(\Lambda^{\rm Narain})$ (\ref{GLambda2}) 
are equal to $\frac1{\sqrt 2}G_{\cal C}$ except for $-\frac 1{\sqrt 2}G_{16} A'^T$.
By the condition (\ref{A'G16^T=0mod2}), $A' {G_{16}}^T\equiv 0$ mod $2$, 
so the components of $\frac1{\sqrt 2}G_{\cal C}$ are ${\sqrt 2}\times$ integers.
Therefore, the $\frac c{\sqrt 2}$ part of Construction A is generated by multiplying 
$G(\Lambda^{\rm Narain})$
by $(w^1,\ldots,w^d)$ and
by $(v^1,\ldots,v^{8})$.
Also, since $-S A'^T$ also has components that are ${\sqrt 2}\times$ integers,
The shifts $(m_{d+1},\ldots,m_{2d})$ and $(m_{2d+9},\ldots,m_{2d+16})$ 
are also realized by $(n_1\,\ldots,n_d)$ and $(v^9,\ldots,v^{16})$.
Furthermore,
the shifts of $(m_{1},\ldots,m_{d})$ and $(m_{2d+1},\ldots,m_{2d+8})$ are also achieved, 
as in the case of $G_{16}$, by multiplying the even $(w^1,\ldots,w^d)$ and $(v^1,\ldots,v^{8})$ 
by $\frac1{\sqrt 2}I_d$ in the top-left block and $\frac1{\sqrt 2}I_8$ in the left half of $\frac1{\sqrt 2}G_{16}$.
Thus we have shown that the heterotic Narain lattice generated 
by $G(\Lambda^{\rm Narain})$ (\ref{GLambda2}) is identical to the code lattice 
constructed by Construction A from the generator matrix $G_{\cal C}$ (\ref{GC}).

\section{Example
}
Let us consider a specific example of the binary code that constitutes 
the heterotic Narain lattice discussed in the previous section.
First, we can choose $\Lambda_{\rm int}$ as one of two even self-dual lattices: 
the $E_8\oplus E_8$ lattice and the $D_{16}^+=D_{16}\bigcup (D_{16}+[1])$ lattice,
corresponding to $E_8\times E_8$ heterotic string and Spin$(32)/\ZZ_2$ heterotic string, respectively,
where $D_{16}$ represents the root lattice and ``$[1]$'' is the highest weight of its spinner representation.
Both can be constructed from a doubly-even self-dual code of length 16 using Construction A (see Appendix).

$E_8\oplus E_8$ corresponds to the direct sum of the extended Hamming code $e_8$; this is decomposable.
The generator matrix is 
\setlength{\tabcolsep}{10pt}
\renewcommand{\arraystretch}{0.6}
\setlength{\arraycolsep}{3pt}
\beqa
G_{e_8\times e_8}&=&
\left(
\begin{array}{ccccccccccc|ccccccccccc}
  1&&&&&&&&&&&&0&1&1&1 &&&&& \\
  &1&&&&&&&&&&&1&0&1&1 &&&&&\\
  &&1&&&&&&&&&&1&1&0&1 &&&&&\\
  &&&1&&&&&&&&&1&1&1&0 &&&&&\\
\rule{0mm}{4mm}  
&&&&&1&&&&&&&&&&&&0&1&1&1  \\
  &&&&&&1&&&&&&&&&&&1&0&1&1 \\
  &&&&&&&1&&&&&&&&&&1&1&0&1 \\
  &&&&&&&&1&&&&&&&&&1&1&1&0 \\  
\end{array}
\right).
\label{GE8xE8text}
\eeqa

On the other hand, the code corresponding to $D_{16}^+$ is an indecomposable code
$d_{16}^+$, which cannot be written as a direct sum, and is defined by the following generating matrix (see Appendix):
\beqa
G_{d_{16}^+}&=&
\left(
\begin{array}{ccccccccccc|cccccccccc}
  1&&&&&&&&&&&&0&1&1&1 &1&1&1&1 \\
  &1&&&&&&&&&&&1&0&1&1 &1&1&1&1\\
  &&1&&&&&&&&&&1&1&0&1 &1&1&1&1\\
  &&&1&&&&&&&&&1&1&1&0 &1&1&1&1\\
\rule{0mm}{4mm}  
&&&&&1&&&&&&&1&1&1&1&0&1&1&1  \\
  &&&&&&1&&&&&&1&1&1&1&1&0&1&1 \\
  &&&&&&&1&&&&&1&1&1&1&1&1&0&1 \\
  &&&&&&&&1&&&&1&1&1&1&1&1&1&0 \\  
\end{array}
\right).
\label{GSpin32/Z2text}
\eeqa
Although $e_8\oplus e_8$ and $d_{16}^+$ are different codes, they 
have the same weight enumerator polynomial, and therefore the corresponding lattice theta functions are also equal.

As the simplest case, consider the $(17, 1)$ Lorentzian even self-dual lattice with $d = 1$.
In the case of bosonic string, when compacted to a one-dimensional torus, 
the Narain lattice is constructed by Construction A from the code generated by a $1\times 2$ matrix $(1 | 0)$.
This corresponds to the case where the $B$ field is $0$.
On the other hand, $A'$ determined by the gauge field $A'=\sqrt{2}A$ is a $16$-component $\FF_2$-vector,
and we can take any row vector of $G_{e_8\times e_8}$ (\ref{GE8xE8text}) or $G_{d_{16}^+}$ (\ref{GSpin32/Z2text})
as $A'$ that satisfies the conditions (\ref{A'G16^T=0mod2})(\ref{A'A'^T=0mod2}).
In this case, $G_{16}=G_{e_8\times e_8}$
or $G_{d_{16}^+}$ is doubly-even, so
$-\frac12 A'{A'}^T=-AA^T$
is $0$ mod $2$.
In all, from the code generated by the $9\times 18$ matrix $G_{\cal C}$:
\setlength{\tabcolsep}{10pt}
\renewcommand{\arraystretch}{0.7}
\setlength{\arraycolsep}{3pt}
\beqa
G_{\cal C}&=&
\left(
\begin{array}{ccc}
1 &0  &  A' \\
 0 & 0 &   G_{16}
\end{array}
\right),
\eeqa
we can construct by Construction A the Narain lattice for the $E_8\times E_8$ 
or Spin$(32)/\ZZ_2$ heterotic string with $G=1$, $B=0$, and $A=\frac{A'}2$.

If we compactify $E_8\times E_8$ heterotic string onto a more general $d$-dimensional torus,
we can take, as $A'$ that satisfies the condition (\ref{A'G16^T=0mod2}), 
what is obtained by left-multiplying $G_{16}$ by any $d\times 8$ matrix $N$ with $\FF_2$ entries.
For example, suppose that we consider compactification to a two-dimensional torus $d=2$ and take 
\beqa
N&=&\left(
\begin{array}{ccccc}
1&1&0&\cdots&0\\
0&1&0&\cdots&0
\end{array}
\right)
\eeqa
as $N$, then we can take  
\beqa
A'&=&NG_{e_8\times e_8}
\n&=&
\left(
\begin{array}{ccccc}
1&1&0&\cdots&0\\
0&1&0&\cdots&0
\end{array}
\right)\left(
\begin{array}{ccccccccccc|ccccccccccc}
  1&&&&&&&&&&&&0&1&1&1 &&&&& \\
  &1&&&&&&&&&&&1&0&1&1 &&&&&\\
  &&1&&&&&&&&&&1&1&0&1 &&&&&\\
  &&&1&&&&&&&&&1&1&1&0 &&&&&\\
\rule{0mm}{4mm}  
&&&&&1&&&&&&&&&&&&0&1&1&1  \\
  &&&&&&1&&&&&&&&&&&1&0&1&1 \\
  &&&&&&&1&&&&&&&&&&1&1&0&1 \\
  &&&&&&&&1&&&&&&&&&1&1&1&0 \\  
\end{array}
\right)
\n&=&
\left(
\begin{array}{cccccccccc|cccccccccccc}
  1&1&0&0 &&0&0&0&0&&&1&1&0&0 &&0&0&0&0 \\
  0&1&0&0 &&0&0&0&0&&&1&0&1&1 &&0&0&0&0
\end{array}
\right)
\eeqa
 that satisfy the conditions. In this case, we have
\beqa
-\frac12 A'{A'}^T
&\equiv&
\left(
\begin{array}{cc}
0&1\\
1&0
\end{array}
\right)~~~\mbox{mod $2$},
\eeqa
resulting in a shift of this amount in the $B$ part of the generator matrix of 
the code that constructs the Narain lattice of bosonic string.

\section{Heterotic Narain lattice constructed from codes over $\FF_3$
}
So far, we have considered constructing a heterotic Narain lattice using a binary code. 
Next let us construct a heterotic Narain lattice using a ternary code, i.e., a code over $\FF_3$, 
using Construction A${}_{\CCsub}$.

In \cite{MizoguchiOikawa}, we showed that for any integer $k$ 
greater than or equal to 2, the lattice constructed 
by ``Construction A${}_{\gsmall}$'' from a code over $\ZZ_k$ 
(or $\FF_k$ if $k$ is a prime number) of length $2n$ 
generated by the generator matrix $(I_n|B')$, 
where $I_n$ is an $n \times n$ identity matrix, and $B'$ is 
an arbitrary $\ZZ_k$-(or $\FF_k$-)valued $n \times n$ antisymmetric matrix, coincides with 
the momentum-wounding lattice of the Narain CFT of bosonic string 
compactified onto a $(k-1)n$-dimensional torus with self-dual radius $R=\sqrt{2}$.
In this case, the metric $g$ is taken to be the tensor product of 
$I_n$ and the inverse matrix $C_{\gsub}^{-1}$ of the Cartan matrix 
of the Lie algebra $\g=SU(k)$, 
and the B field $B$ is taken to be the tensor product of 
$B'$ and $C_{\gsub}^{-1}$.
This lattice construction  ``Construction A${}_{\gsmall}$" 
is ultimately nothing more than constructing a lattice by 
``gluing \cite{ConwaySloan}" the root lattices of a Lie algebra $\g$, 
whose ``dual quotient \footnote{For a Lie algebra $\gsmall$, 
let  $\Lambda^{\gsub}_R$ be the root lattice of $\gsmall$ and 
$\Lambda^{\gsub}_W$ be the weight lattice, then the coset module 
$\Lambda^{\gsub}_W/\Lambda^{\gsub}_R$ is called the ``dual quotient'' 
(of $\Lambda^{\gsub}_R$) \cite{ConwaySloan}.}" is $k$, 
with a given code over $\ZZ_k$ (or $\FF_k$ if $k$ is prime) 
as the ``glue code".
With $g$ and $B$ as above, 
the winding lattice is a direct sum of the weight lattices 
$\Lambda^{\gsmall}_W$ of $\g$, 
and the momentum lattice is a direct sum of the root lattices 
$\Lambda^{\gsmall}_R$ of $\g$.
The coset obtained by dividing the former by the latter can be identified 
as the finite field or ring in which the code takes values.
The construction of lattices from codes based on such correspondences 
is a natural extension of Construction A and Construction A${}_{\CCsub}$.

In the original Construction A${}_{\CCsub}$, 
given a code ${\cal C}$ over $\FF_3$, 
we associated a lattice 
\beqa
\Lambda_{\cal C}&:=&\left\{
\left.
\frac{c+(\omega-\bar\omega)m}{\sqrt{3}}
\right|\,c\in{\cal C},~~m\in{\cal E}^{2n}
\right\},
\eeqa
where ${\cal E}$ denotes the Eisenstein integers
\beqa
\cal E&:=&\{
m_1+m_2\,\omega\,|\,m_1,m_2\in\ZZ
\}.
\eeqa
In \cite{MizoguchiOikawa}, 
using the fact that the dual quotient of the $SU(3)$ root lattice is $3$, 
the construction of an equivalent lattice
\beqa
\Lambda_{\cal C}&:=&\left\{
c\,\omega^{SU(3)}_1+m
\in (\Lambda_W^{SU(3)})^{2n}
\left|\,
c\in{\cal C},~m\in (\Lambda_R^{SU(3)})^{2n}
\right.
\right\},
\label{LambdaCg}
\eeqa
was called Construction A${}_{\gsmall}$ (where $\g=SU(3)$).
This procedure is exactly the glueing of $SU(3)$ root lattices 
with the code ${\cal C}$ as the glue code.

In order for the code lattice constructed in this way  and the Narain lattice to coincide, 
we first compactify heterotic string to a direct product of $n$ $SU(3)$ dual tori of radius $R=\sqrt 2$, 
following \cite{MizoguchiOikawa}.
Let $R=\sqrt 2$ in (\ref{matrix_representation}) and take the generator matrix 
of the momentum-winding lattice $\Lambda^{\rm Narain}$ as \footnote{In this paper, 
we change the matrix representation of tensor products in metrics, vielbeins, etc. 
from \cite{MizoguchiOikawa} so that they become block diagonal.}
\setlength{\tabcolsep}{10pt}
\renewcommand{\arraystretch}{1.0}
\setlength{\arraycolsep}{5pt}
\beqa
G(\Lambda^{\rm Narain})&:=&
\left(
\begin{array}{ccc}
 \gamma & (B^T-A\,A^T)\gamma^{T-1}    &  \sqrt 2 A \\
 0 & \gamma^{T-1}  &  0 \\
 0 & -\sqrt{2}\gamma_{16} A^T \gamma^{T-1} &   \gamma_{16}
\end{array}
\right),
\label{GLambdaF3}
\eeqa
%\beqa
%g_{ij}=\left(
%\begin{array}{cccc}
%%
%\framebox{\scriptsize 
%$\begin{array}{cc}
%\frac23&\frac13\\
%\frac13&\frac23
%\end{array}
%$}
%%
%&&&\\
%&%
%\framebox{\scriptsize 
%$\begin{array}{cc}
%\frac23&\frac13\\
%\frac13&\frac23
%\end{array}
%$}
%%
%&&
%\\
%&&\ddots&\\
%&&&
%\framebox{\scriptsize 
%$\begin{array}{cc}
%\frac23&\frac13\\
%\frac13&\frac23
%\end{array}
%$}
%\end{array}
%\right)
%&=&
%I_n\otimes C_{SU(3)}^{-1}
%\label{gijF3}
%\eeqa
\setlength{\tabcolsep}{10pt}
\renewcommand{\arraystretch}{0.6}
\setlength{\arraycolsep}{3pt}
\beqa
\gamma&=&
\left(
\begin{array}{cccc}
  \gamma_W&&&   \\
  &\gamma_W&&   \\
  &   &\ddots&\\
  &&&\gamma_W   
\end{array}
\right)~=~
%I_n\otimes \gamma_W,
\gamma_W\otimes I_n,
\label{gammaF3}
\eeqa
\beqa
\gamma^{T-1}&=&
\left(
\begin{array}{cccc}
  \gamma_W^{T-1}&&&   \\
  &\gamma_W^{T-1}&&   \\
  &   &\ddots&\\
  &&&\gamma_W^{T-1}
\end{array}
\right)
 ~=~
%I_n\otimes \gamma_W^{T-1},
\gamma_W^{T-1}\otimes I_n,
\label{gammaT-1F3}
\eeqa
\beqa
\gamma_W&=&
\frac1{\sqrt 6}\left(
\begin{array}{cc}
  \sqrt 3& 1  \\
  \sqrt 3&-1  
\end{array}
\right)
~=~
\left(
\begin{array}{c}
  \omega_1^{SU(3)}  \\
  \omega_2^{SU(3)}  
\end{array}
\right),\n
\gamma_W^{T-1}&=&
\frac1{\sqrt 2}\left(
\begin{array}{cc}
  1& \sqrt 3  \\
  1&-\sqrt 3  
\end{array}
\right)
~=~
\left(
\begin{array}{c}
  \alpha_1^{SU(3)}  \\
  \alpha_2^{SU(3)}  
\end{array}
\right).
\eeqa
\beqa
B_{ij}&=&(C_{SU(3)}^{-1}\otimes B')_{ij}~=~
(\gamma_W \gamma_W^T\otimes B')_{ij},
\label{BijF3}
\eeqa
using the $\FF_3$-valued $n\times n$ antisymmetric matrix $B'$.
On the other hand, the gauge field in this case is
\beqa
A_i^{~I}&=&(\gamma_W\otimes \sqrt 2A')_i^{~I}.
\label{AiIF3}
\eeqa
The factor $\sqrt 2$ of $A'$ is different from the factor $\frac1{\sqrt 2}$ in the binary code case. 
The reason for this is that if it were the same as in the binary code case, 
$A'{A'}^T$ would be required to be divisible by $2$, 
which would be inconsistent with the definition of $A'$ mod $3$.

Now, under the above setting, consider a code over $\FF_3$ 
generated by a $(n+4)\times(2n+8)$ matrix :
\setlength{\tabcolsep}{10pt}
\renewcommand{\arraystretch}{0.8}
\setlength{\arraycolsep}{5pt}
\beqa
G_{\cal C}&=&
\left(
\begin{array}{ccc}
I_n &{B'}^T\!\!-2A'{A'}^T  &  2A' \\
 0 & 0 &   G_{8}
\end{array}
\right),
\label{GCF3}
\eeqa
whose entries take values in $\FF_3$.
$G_{8}$ is a direct sum of tetra codes :
%?\setlength{\tabcolsep}{10pt}
\renewcommand{\arraystretch}{0.7}
\setlength{\arraycolsep}{3pt}
\beqa
G_8&=&\left(
\begin{array}{ccccc|ccrcr}
  1&~0~&0&~0~&&&1&1\,&\,0\,&\!0  \\
  0&1&0&0&&&1&-1\,&0&\!0 \\
  0&0&1&0&&&0&0\, &1&\!1 \\
  0&0&0&1&&&0&0\,&1&\!-1
\end{array}
\right).
\eeqa
This is the generator matrix for the code that realizes the $E_8\oplus E_8$ lattice 
as $\Lambda_{\rm int}$ by construction A${}_{\CCsub}$ \footnote{$D_{16}^+$ 
does not include $A_2^{\oplus 8}$ as a sublattice, 
so it cannot be realized (at least not in this way) from the code over $F_3$.}.
For this $G_{\cal C}$ to construct a Narain lattice of $E_8\times E_8$ heterotic string, 
$G_{\cal C}$ must be self-dual, i.e., $G_{\cal C}G_{\cal C}^T\equiv 0$ 
mod $3$ \footnote{In the case of the codes over $\FFsmall_3$ 
we are considering, there is no need for a condition similar to 
the doubly-even condition imposed for binary codes. 
This condition is required for modular T-invariance of the lattice theta, 
but in this case, the square of the lengths of the base weights $\omega_1^{SU(3)}$ 
and $\omega_2^{SU(3)}$ is $\frac23$, so if the code is self-dual, 
it automatically becomes modular T-invariant.}.
Since $G_8 G_8^T\equiv 0$ mod $3$, the only condition for this to be the case is
\beqa
A'G_8^T\equiv 0~~~\mbox{mod $3$},
\eeqa
which we require.
As $A'$ that satisfies this, let us take $A'=N G_8$, 
where $N$ is again an arbitrary $n\times 4$ integer matrix.
Then $A'A'^T\equiv 0$ mod $3$, which disappears from $G_{\cal C}$ (\ref{GCF3}).

The fact that the code generated from $G_8$ constructs 
the $E_8\oplus E_8$ lattice by Construction A${}_{\CCsub}$ means that 
the generator matrix $\gamma_{16}$ for which $\Lambda_{\rm int}$ is 
the $E_8\oplus E_8$ lattice can be written as
\setlength{\tabcolsep}{10pt}
\renewcommand{\arraystretch}{0.8}
\setlength{\arraycolsep}{2pt}
\beqa
\gamma_{16}&=&\left(
\begin{array}{l}
~~\gamma_W\otimes G_{8}\\
%S
%
%
\gamma_W^{T-1}\otimes
\left(\begin{array}{cc}
0&~I_4
\end{array}\right)
\end{array}
\right),
%,~~~S:=(O\otimes \gamma_W^{T-1}~~ I_4\otimes \gamma_W^{T-1}),
%
\label{gamma16F3}
\eeqa
that is,
\setlength{\tabcolsep}{10pt}
\renewcommand{\arraystretch}{0.7}
\setlength{\arraycolsep}{2pt}
\beqa
\gamma_{16}&=&\left(
\begin{array}{rcr}
\gamma_W \otimes 
\left(
\begin{array}{cccc}
1&~0~&0&~0 \\
  0&1&0&0 \\
  0&0&1&0 \\
  0&0&0&1
\end{array}
\right)
&&
\gamma_W\otimes \left(
\begin{array}{cccc}
1&1\,&\,0\,&\!0  \\
1&-1\,&0&\!0 \\
0&0\, &1&\!1 \\
0&0\,&1&\!-1
\end{array}
\right)
\\
\\
\gamma_W^{T-1}\otimes 
\left(
\begin{array}{cccc}
0&~0~&0&~0 \\
  0&0&0&0 \\
  0&0&0&0 \\
  0&0&0&0
\end{array}
\right)
&&
\gamma_W^{T-1}\otimes\left(
\begin{array}{cccc}
1&~0~&0&~0 \\
  0&1&0&0 \\
  0&0&1&0 \\
  0&0&0&1
  \end{array}
\right)
\end{array}
\right).
\label{gamma16F3'}
\eeqa
This is a generalization of (\ref{gamma16}) (Conway-Sloan\cite{ConwaySloan}
Chapter 7 p.183).
Indeed, let 
\beqa
\gamma_{E_8}&:=&
\left(
\begin{array}{rcr}
\gamma_W\otimes \left(
\begin{array}{cc}
1&~0~ \\
  0&1 \\
\end{array}
\right)
&&
\gamma_W\otimes \left(
\begin{array}{cc}
1&1\,\\
1&-1\,
\end{array}
\right)
\\
\\
\gamma_W^{T-1}\otimes \left(
\begin{array}{cc}
0&~0~ \\
  0&0 
\end{array}
\right)
&&
\gamma_W^{T-1}\otimes \left(
\begin{array}{cc}
1&~0~ \\
  0&1 
    \end{array}
\right)
\end{array}
\right)
\eeqa
be one direct sum component of (\ref{gamma16F3'}), 
then $\gamma_{E_8}\gamma_{E_8}^T$ is integral unimodular and 
can be transformed to the $E_8$ Cartan matrix by multiplying it 
on both sides by some integer matrix and its transpose.

Now, let us show that the code lattice $\Lambda_{{\cal C}_{G_{\cal C}}}$ 
constructed from $G_{\cal C}$ (\ref{GCF3}) by Construction A${}_{\CCsub}$ 
(\ref{LambdaCg}) is identical to the lattice generated by $G(\Lambda^{\rm Narain})$ (\ref{GLambdaF3}).
First, by substituting (\ref{gammaF3})(\ref{gammaT-1F3})(\ref{BijF3})(\ref{AiIF3})(\ref{gamma16F3}) 
into $G(\Lambda^{\rm Narain})$ (\ref{GLambdaF3}), we get 
\setlength{\tabcolsep}{10pt}
\renewcommand{\arraystretch}{1}
\setlength{\arraycolsep}{5pt}
\beqa
G(\Lambda^{\rm Narain})&:=&
\left(
\begin{array}{ccc}
\gamma_W\otimes I_n &\gamma_W\otimes  ({B'}^T\!\!-A'{A'}^T)    &  \gamma_W\otimes 2 A' \\
 0 & \gamma_W^{T-1}\otimes I_n  &  0 \\
 0 & \gamma_W\otimes(-2G_8{A'}^T) &  \gamma_W\otimes G_8\\
 0 & \gamma_W^{T-1}\otimes(-2(0~~I_4){A'}^T) &  \gamma_W^{T-1}\otimes (0~~I_4)
\end{array}
\right).
\label{GLambdaF3'}
\eeqa
Then all we need to do is show that $(x_1,\ldots,x_{n+4}~;~m_1^1,\ldots,m_{2n+8}^2)\in\FF_3^{n+4}\times\ZZ^{4n+16}$
that satisfies 
\setlength{\tabcolsep}{10pt}
\renewcommand{\arraystretch}{0.5}
\setlength{\arraycolsep}{1pt}
\beqa
&&G_{\cal C}^T
\left(
\begin{array}{cc}
x_1&0\\ 
\vdots&\vdots\\
x_n&0\\ 
\rule{0mm}{5mm}
x_{n+1}&0\\
\vdots&\vdots\\
x_{n+4}&0\\ 
\end{array}
\right)\gamma_W
+
\left(
\begin{array}{cc}
m_1^{~1}&m_1^{~2}\\ 
\vdots&\vdots\\
m_n^{~1}&m_n^{~2}\\ 
\rule{0mm}{6mm}
m_{n+1}^{~1}&m_{n+1}^{~2}\\ 
\vdots&\vdots\\
m_{2n}^{~1}&m_{2n}^{~2}\\ 
\rule{0mm}{6mm}
m_{2n+1}^{~1}&m_{2n+1}^{~2}\\ 
\vdots&\vdots\\
m_{2n+4}^{~1}&m_{2n+4}^{~2}\\ 
\rule{0mm}{6mm}
m_{2n+5}^{~1}&m_{2n+5}^{~2}\\ 
\vdots&\vdots\\
m_{2n+8}^{~1}&m_{2n+8}^{~2}
\end{array}
\right)
\gamma_W^{T-1}\n
&=&((w_1^{~1}w_1^{~2}\cdots w_n^{~1}w_n^{~2}~
n_1^{~1}n_1^{~2}\cdots n_n^{~1}n_n^{~2}~\n
&&w_{n+1}^{~1}w_{n+1}^{~2}\cdots w_{n+4}^{~1}w_{n+4}^{~2}~
n_{n+1}^{~1}n_{n+1}^{~2}\cdots n_{n+4}^{~1}n_{n+4}^{~2})
G(\Lambda^{\rm Narain}))^{1\otimes T},
\eeqa
where $1\otimes T$ is the transpose operation only in the vector space of the code,
and 
$(w_1^1,\ldots,w_{n+4}^2~;~n_1^1,\ldots,n_{n+4}^2)\in\ZZ^{4n+16}$
 correspond one to one. This can be verified elementarily.

\section{Heterotic Narain lattice constructed from codes over $\FF_5$
}
The $E_8$ lattice can be constructed from a code over $\FF_5$ of length 2 
by glueing together the $SU(5)$ root lattices \cite{Ebeling}, 
so it can also be used to construct 
a heterotic Narain lattice from codes over $\FF_5$.
As in previous examples, consider a code over $\FF_5$ generated 
by the $(n+2)\times(2n+4)$ matrix 
\setlength{\tabcolsep}{10pt}
\renewcommand{\arraystretch}{0.8}
\setlength{\arraycolsep}{5pt}
\beqa
G_{\cal C}&=&
\left(
\begin{array}{ccc}
I_n &{B'}^T\!\!-2A'{A'}^T  &  2A' \\
 0 & 0 &   G_{4}
\end{array}
\right)
\label{GCF5}
\eeqa
with entries taking values in $\FF_5$, where
\setlength{\tabcolsep}{10pt}
\renewcommand{\arraystretch}{0.7}
\setlength{\arraycolsep}{3pt}
\beqa
G_4&=&\left(
\begin{array}{ccc|ccc}
  1&0&&&2&0  \\
  0&1&&&0&2 \\
\end{array}
\right).
\eeqa
The fact that the code generated from this $G_4$
constructs the $E_8\oplus E_8$ lattice by Construction A${}_{\gsmall}$ with $\g=SU(5)$,
that is, by glueing together the $SU(5)$ root lattices with a code over $\FF_5$ as the glue code, 
means that the generator matrix $\gamma_{16}$ of the $E_8\oplus E_8$ lattice can now be written as 
\setlength{\tabcolsep}{10pt}
\renewcommand{\arraystretch}{0.8}
\setlength{\arraycolsep}{2pt}
\beqa
\gamma_{16}&=&\left(
\begin{array}{l}
~~\gamma_W^{SU(5)}\otimes G_{4}\\
\gamma_W^{SU(5)\,T-1}\otimes
\left(\begin{array}{cc}
0&~I_2
\end{array}\right)
\end{array}
\right),
\label{gamma16F5}
\eeqa
\beqa
\gamma_W^{SU(5)}&=&
\left(
\begin{array}{c}
  \omega_1^{SU(5)}  \\
  \vdots\\
  \omega_4^{SU(5)}  
\end{array}
\right),~~
\gamma_W^{SU(5)\,T-1}~=~
\left(
\begin{array}{c}
  \alpha_1^{SU(5)}  \\
  \vdots\\
  \alpha_4^{SU(5)}  
\end{array}
\right).
\eeqa
Indeed, let 
\beqa
\gamma_{E_8}^{\FFsub_5}=\left(
\begin{array}{cc}
\gamma_W^{SU(5)}&2\gamma_W^{SU(5)}\\
0&\gamma_W^{SU(5)\,T-1}
\end{array}
\right)
\eeqa
be one of the direct sum components, then 
$\gamma_{E_8}^{\FFsub_5}\gamma_{E_8}^{\FFsub_5\,T}$
is again integral unimodular, 
and can be multiplied by an integer matrix and its transpose on both sides 
to form the $E_8$ Cartan matrix.

Since $G_{\cal C}$ (\ref{GCF5}) is self-dual, we require
\beqa
A'G_4^T&\equiv& 0~~~\mbox{mod $5$}.
\eeqa
Such $A'$ can again be taken as $A'=N G_4$, where $N$ is any $n\times 2$ integer matrix.
On the other hand, the generator matrix of the momentum-winding lattice 
$\Lambda^{\rm Narain}$ is
\setlength{\tabcolsep}{10pt}
\renewcommand{\arraystretch}{1.0}
\setlength{\arraycolsep}{5pt}
\beqa
G(\Lambda^{\rm Narain})&:=&
\left(
\begin{array}{ccc}
 \gamma & (B^T-A\,A^T)\gamma^{T-1}    &  \sqrt 2 A \\
 0 & \gamma^{T-1}  &  0 \\
 0 & -\sqrt{2}\gamma_{8} A^T \gamma^{T-1} &   \gamma_{16}
\end{array}
\right),
\label{GLambdaF5}
\eeqa
\setlength{\tabcolsep}{10pt}
\renewcommand{\arraystretch}{0.7}
\setlength{\arraycolsep}{3pt}
\beqa
\gamma&=&
\gamma_W^{SU(5)}\otimes I_n,~~~
%\label{gammaF5}\\
\gamma^{T-1}~=~
\gamma_W^{SU(5)\;T-1}\otimes I_n,
\label{gammaT-1F5}
\eeqa
\beqa
B_{ij}&=&(C_{SU(5)}^{-1}\otimes B')_{ij}~=~
(\gamma_W^{SU(5)} \gamma_W^{SU(5)\;T}\otimes B')_{ij},
\label{BijF5}\\
A_i^{~I}&=&(\gamma_W^{SU(5)}\otimes \sqrt 2A')_i^{~I}.
\label{AiIF5}
\eeqa
Then, as in the case of $\FF_3$, we can show that 
$(x_1,\ldots,x_{n+2}~;~m_1^1,\ldots,m_{2n+4}^4)\in\FF_5^{n+2}\times\ZZ^{8n+16}$
satisfying
\setlength{\tabcolsep}{10pt}
\renewcommand{\arraystretch}{0.5}
\setlength{\arraycolsep}{1pt}
\beqa
&&G_{\cal C}^T
\left(
\begin{array}{cccc}
x_1&0&0&0\\ 
\vdots&&\vdots&\\
x_n&0&0&0\\ 
\rule{0mm}{5mm}
x_{n+1}&0&0&0\\
x_{n+2}&0&0&0\\ 
\end{array}
\right)\gamma_W^{SU(5)}
+
\left(
\begin{array}{ccc}
m_1^{~1}&\cdots&m_1^{~4}\\ 
\vdots&\vdots&\vdots\\
m_n^{~1}&\cdots&m_n^{~4}\\ 
\rule{0mm}{6mm}
m_{n+1}^{~1}&\cdots&m_{n+1}^{~4}\\ 
\vdots&\vdots&\vdots\\
m_{2n}^{~1}&\cdots&m_{2n}^{~4}\\ 
\rule{0mm}{6mm}
m_{2n+1}^{~1}&\cdots&m_{2n+1}^{~4}\\ 
m_{2n+2}^{~1}&\cdots&m_{2n+2}^{~4}\\ 
\rule{0mm}{6mm}
m_{2n+3}^{~1}&\cdots&m_{2n+3}^{~4}\\ 
m_{2n+4}^{~1}&\cdots&m_{2n+4}^{~4}
\end{array}
\right)
\gamma_W^{SU(5)\;T-1}\n
&=&((w_1^{~1}\cdots w_1^{~4}~\cdots~w_n^{~1}\cdots w_n^{~4}
n_1^{~1}\cdots n_1^{~4}~\cdots~n_n^{~1}\cdots n_n^{~4}\n
&&w_{n+1}^{~1}\cdots w_{n+1}^{~4} w_{n+2}^{~1}\cdots w_{n+2}^{~4}~
n_{n+1}^{~1}\cdots n_{n+1}^{~4} n_{n+2}^{~1}\cdots n_{n+2}^{~4})
G(\Lambda^{\rm Narain}))^{1\otimes T}
\eeqa
and $(w_1^1,\ldots,w_{n+2}^4~;~n_1^1,\ldots,n_{n+2}^4)\in\ZZ^{8n+16}$
correspond one to one. This shows that the coincidence between 
the code lattice $\Lambda_{{\cal C}_{G_{\cal C}}}$ constructed from 
$G_{\cal C}$ (\ref{GCF5}) and the lattice generated by 
$G(\Lambda^{\rm Narain})$ (\ref{GLambdaF5}).

\section{Conclusion and discussion
}
In this paper, we have constructed the Narain lattice, 
which arises from the torus compactification 
of heterotic string, from error correcting codes.
We have first given, in both $E_8\times E_8$ 
and Spin$(32)/\ZZ_2$ heterotic strings, a pair of a binary code and a set of the corresponding metric, B field, 
and background gauge field, such that the lattice constructed from the binary code
by Construction A coincides with the Narain lattice.
We have also given a pair of code over $\FF_3$ or $\FF_5$ 
and a set of corresponding metric, 
B field and background gauge field such that the code lattice constructed by 
Construction A${}_{\gsmall}$ with $\g=SU(3)$ or $SU(5)$ 
from a code over $\FF_3$ or $\FF_5$, respectively, 
coincides with the Narain lattice of $E_8\times E_8$ heterotic string on that background.

As mentioned in the text, Construction A${}_{\gsmall}$ is a gluing of 
root lattices of a Lie algebra such that the finite field or ring over which 
a given code is defined is the dual quotient, 
so we can also consider other Lie algebras and codes over their dual quotients, 
as long as their rank does not exceed the spacetime dimension of heterotic string.
Also, in the case of $E_8\times E_8$ string, 
it is possible to construct the internal $E_8\times E_8$ lattice by glueing  
root lattices of different Lie algebras with the same dual quotient 
(e.g., $SU(2)$ and $E_7$, $SU(3)$ and $E_6$, etc.).

In the previously known correspondence 
between Narain CFTs of the spacetime signature $(d, d)$
and {\em quantum} error correcting codes,
the left half of $G_{\cal C}$ corresponds to the number of Pauli $X$s 
contained in the stabilizer, and the left half corresponds to the number of Pauli $Z$s.
In the case of $(d + 16, d)$ considered in this paper, $16$ symbols are added, 
making it difficult to interpret as it is. This is a topic for future work.

It is unclear at this point how the relationship between classical 
error correcting  
codes and the compactification of superstring theory, as considered in this paper, 
will play a role in modern quantum information technology. 
However, a deeper understanding of the relationship between the two may 
lead to unexpected developments in the future.

\appendix
\section{Codes constructing the $D_{16}^+$ lattice and NSR fermions
}
As is well known, the $E_8$ root lattice, the only eight-dimensional 
Euclidean even self-dual lattice, is constructed from the extended Hamming code $e_8$ by Construction A.
Furthermore, of the two 16-dimensional Euclidean even self-dual lattices, the $E_8\oplus E_8$ lattice 
is trivially constructed from two independent extended Hamming codes, 
and the codes that construct the $D_{16}^+$ lattice are also known.
Since these lattices are associated with modular invariant partition functions of 
16 left-moving fermions of two heterotic strings, 
the codes constructing these lattices by Construction A are naturally related to NSR fermions.
Here we will discuss these relationships.

\subsection{Extended Hamming code $e_8$ to the $E_8$ lattice
}
First, we will revisit how the theta functions of the $E_8$ lattice are constructed from the extended Hamming code $e_8$ from the perspective of NSR fermions.
The generator matrix $G_{e_8}$ of the Extended Hamming code $e_8$ can be taken, by appropriately rearranging the rows and columns, in ``systematic form'':
\setlength{\tabcolsep}{10pt}
\renewcommand{\arraystretch}{0.6}
\setlength{\arraycolsep}{3pt}
\beqa
G_{e_8}&=&
\left(
\begin{array}{ccccc|ccccc}
  1&0&0&0&&&0&1&1&1  \\
  0&1&0&0&&&1&0&1&1 \\
  0&0&1&0&&&1&1&0&1 \\
  0&0&0&1&&&1&1&1&0 \\
\end{array}
\right),
\eeqa
in which the left half of the matrix is the identity matrix $I_4$.
The right half of this $G_{e_8}$ is a $4\times 4$ matrix, 
which is obtained by adding to it (in $\FF_2$) 
a matrix whose elements are all $1$ :
\beqa
\left(
\begin{array}{ccccccccc}
  1&1&1&1  \\
  1&1&1&1  \\
  1&1&1&1  \\
  1&1&1&1  
\end{array}
\right)
\label{Z2}
\eeqa
and inverting $0$ to $1$ and vice versa.
We call this the ``$\ZZ_2$ inversion."
The extended Hamming code $e_8$ consists of all $c\in\FF_2^8$ obtained by multiplying $G$ by $x=(x_1\;x_2\;x_3\;x_4)\in\FF_2^4$ from the left:
\beqa
x\,G=c=(c_1\;c_2\;c_3\;c_4\;c_5\;c_6\;c_7\;c_8).
\eeqa

From this code, the theta function of the lattice 
$\Lambda_{e_8}$ constructed by Construction A is 
given by the sum of the products of eight level-$1$ 
theta functions over all codewords:
\beqa
\sum_{c\in e_8}\prod_{i=1}^4(\Theta_{c_i,1}\Theta_{c_{i+4},1}),
\label{e8codelatticetheta}
\eeqa
where a level-$k$ theta function is defined as 
\beqa
\Theta_{m,k}(\tau,z)&:=&\sum_{n\in \ZZsub} q^{k(n+\frac m{2k})^2} e^{2\pi i (n+\frac m{2k})z},~~~q:=e^{2\pi i \tau},
\eeqa
and $\Theta_{c_i,1}(\tau,0)=:\Theta_{c_i,1}$.

If we are only calculating the lattice theta function, 
the order of multiplication of the level-$1$ theta 
corresponding to each symbol in the code does not matter;
by freely changing the order of multiplication, due to the fact that 
$\Theta_{0,1}=\vartheta_3(2\tau)$, $\Theta_{1,1}=\vartheta_2(2\tau)$,
it reduces to the well-known formula for 
finding the lattice theta from the weight enumerator.
Instead, we fix a specific pair of symbols in the code 
as in (\ref{e8codelatticetheta}), 
and calculate the product without changing the order.
Then, by using the composition formula of the theta function, we have
 \beqa
 \Theta_{0,1}^2&=&\Theta_{0,2}^2+\Theta_{2,2}^2~=~\frac12(\vartheta_3^2+\vartheta_4^2),\n
 \Theta_{1,1}^2&=&\Theta_{2,2}\Theta_{0,2}+\Theta_{0,2}\Theta_{2,2}~=~\frac12(\vartheta_3^2-\vartheta_4^2)
 \label{NSJacobi}
 \eeqa
 \beqa
 \Theta_{0,1}\Theta_{1,1}&=&\Theta_{1,2}^2+\Theta_{-1,2}^2~=~\frac12 (\vartheta_2^2+\vartheta_1^2)\n
 \Theta_{1,1}\Theta_{0,1}&=&\Theta_{1,2}\Theta_{-1,2}+\Theta_{-1,2}\Theta_{1,2}~=~\frac12 (\vartheta_2^2-\vartheta_1^2),
  \label{RJacobi}
 \eeqa
which allows us to distinguish between the NS sector and the R sector 
depending on whether it is $(\Theta_{0,1}^2)(\Theta_{1,1}^2)$
or $(\Theta_{0,1}\Theta_{1,1})(\Theta_{1,1}\Theta_{0,1})$,
even if the contribution to the lattice theta is the same.

Thus, by considering the sum of products of fixed level-1 theta pairs,
it can be easily shown that the extended Hamming code $e_8$ 
constructs the $E_8$ lattice as follows:
\begin{itemize}
\item{The case $\sum_{i=1}^4 x_i\equiv0$ mod $2$
\\
In this case, since (\ref{Z2}) becomes $0$ mod $2$ when added even times, 
the code $c$ simply becomes 
\beqa
(x_1\;x_2\;x_3\;x_4~~x_1\;x_2\;x_3\;x_4).
\eeqa
Therefore, the contribution of this code to the theta function 
associated with that lattice is
\beqa
\Theta_{x_1,1}^2\Theta_{x_2,1}^2\Theta_{x_3,1}^2\Theta_{x_4,1}^2.
\eeqa
Due to (\ref{NSJacobi}), each $\Theta^2$ factor 
correspond to two NS fermions, 
with $x_1=0$ corresponding to the even fermion number sector 
and $x_1=1$ corresponding to the odd fermion number sector.
In addition, since 
$\sum_{i=1}^4 x_i\equiv0$ mod $2$,
the whole thing is a projection onto the even fermion number sector.
Adding these contributions for all $x_i$ that satisfy the condition gives 
\beqa
\sum_{\sum_{i=1}^4 x_i\equiv0~\mbox{\scriptsize mod $2$}}
\Theta_{x_1,1}^2\Theta_{x_2,1}^2\Theta_{x_3,1}^2\Theta_{x_4,1}^2
&=&( \Theta_{0,1}^2)^4+6( \Theta_{0,1}^2)^2( \Theta_{1,1}^2)^2+(( \Theta_{1,1}^2)^2)^4\n
&=&\frac12\left(
(( \Theta_{0,1}^2)^2+( \Theta_{1,1}^2)^2)^4+(( \Theta_{0,1}^2)^2-( \Theta_{1,1}^2)^2)^4
\right)\n
&=&\frac12\left(
(\vartheta_3^2)^4+(\vartheta_4^2)^4
\right)\n
&=&\frac12\left(
\vartheta_3^8+\vartheta_4^8
\right).
\label{E8NS}
\eeqa
}

\item{The case  $\sum_{i=1}^4 x_i\equiv 1$ mod $2$
\\
In this case, the code $c$ is in the form

\beqa
(x_1\;x_2\;x_3\;x_4~~x_1+1\;\;x_2+1\;\;x_3+1\;\;x_4+1)
\eeqa
mod $2$. The contribution of the code of this form to the lattice theta 
is 
\beqa
(\Theta_{x_1,1}\Theta_{x_1+1,1})
(\Theta_{x_2,1}\Theta_{x_2+1,1})
(\Theta_{x_3,1}\Theta_{x_3+1,1})
(\Theta_{x_4,1}\Theta_{x_4+1,1}).
\label{Rsector_to_Theta}
\eeqa
All of these factors are either $(\Theta_{0,1}\Theta_{1,1})$ 
or $(\Theta_{1,1}\Theta_{0,1})$.
According to (\ref{RJacobi}), these in turn correspond to Ramond fermions.
In the usual convention of the theta functions, 
$\vartheta_1=0$ when $z$ or $\nu=0$. 
However, by including this $\vartheta_1$, 
we can distinguish the chirality of the Ramond fermion.
This will become important when we consider the triality of $SO(8)$.
For all $x_i$ that satisfy the condition, adding (\ref{Rsector_to_Theta}) gives
\beqa
&&\sum_{\sum_{i=1}^4 x_i\equiv1~\mbox{\scriptsize mod $2$}}
(\Theta_{x_1,1}\Theta_{x_1+1,1})
(\Theta_{x_2,1}\Theta_{x_2+1,1})
(\Theta_{x_3,1}\Theta_{x_3+1,1})
(\Theta_{x_4,1}\Theta_{x_4+1,1})\n
&=&4(\Theta_{1,1}\Theta_{0,1})(\Theta_{0,1}\Theta_{1,1})^3
+4(\Theta_{1,1}\Theta_{0,1})^3(\Theta_{0,1}\Theta_{1,1})
\n
&=&\frac12
(\vartheta_2^8-\vartheta_1^8).
\label{E8R}
\eeqa
}
\end{itemize}
From the above, the lattice theta constructed from 
the extended Hamming code by Construction A is shown to be, 
with $\vartheta_1=0$, 
\beqa
(\ref{E8NS})+(\ref{E8R})&=&
\frac12\left(
\vartheta_3^8+\vartheta_4^8+\vartheta_2^8
\right),
\label{E8theta}
\eeqa
which is exactly the $E_8$ theta.

It should be emphasized here that our goal here is not 
to extract the lattice theta of the code lattice from 
the extended Hamming code, 
but rather to clarify the structure of the code lattice itself 
by fixing the order of theta multiplications. 
As we will see next, different lattices exist even if the lattice theta is the same.

\subsection{Binary codes constructing the $D_{16}^+$ lattice
}
\label{section:Spin(32)/Z2code}
In this way, by taking the generator matrix of the extended Hamming code 
with an explicit $\ZZ_2$ inversion structure, 
we have clarified the relationship between the lattice constructed 
by Construction A from it and the NSR fermion CFT lattice 
yielding the partition function of heterotic string.
In particular, we have seen that the theta function associated 
with the lattice constructed from such a generator matrix is 
automatically GSO-projected.

In heterotic string, for 16 complex fermions, 
we are able to obtain the partition function of the $E_8\times E_8$ theory 
by performing GSO projection independently on two groups of 
8 complex fermions each.
The lattice that yields its theta function part, i.e., the fermion partition function multiplied by $|\eta(\tau)|^{16}$, as the lattice theta 
is obviously constructed from the code generated by the generator matrix 
\beqa
G_{e_8\times e_8}&=&
\left(
\begin{array}{cccccccccccccccccccc}
  1&&&&&&&&&&0&1&1&1 &&&&& \\
  &1&&&&&&&&&1&0&1&1 &&&&&\\
  &&1&&&&&&&&1&1&0&1 &&&&&\\
  &&&1&&&&&&&1&1&1&0 &&&&&\\
\rule{0mm}{4mm}  
&&&&&1&&&&&&&&&&0&1&1&1  \\
  &&&&&&1&&&&&&&&&1&0&1&1 \\
  &&&&&&&1&&&&&&&&1&1&0&1 \\
  &&&&&&&&1&&&&&&&1&1&1&0 \\  
\end{array}
\right),
\eeqa
where all matrix elements that are not written are $0$.
On the other hand, if we perform GSO on all 16 complex fermions 
simultaneously, we obtain the Spin$(32)/\ZZ_2$
heterotic theory. So we consider an $8\times 16$ generator matrix,
where the left half is the identity matrix and the right half is its $\ZZ_2$ inversion:
\beqa
G_{d_{16}^+}&=&
\left(
\begin{array}{ccccccccccc|cccccccccc}
  1&&&&&&&&&&&&0&1&1&1 &1&1&1&1 \\
  &1&&&&&&&&&&&1&0&1&1 &1&1&1&1\\
  &&1&&&&&&&&&&1&1&0&1 &1&1&1&1\\
  &&&1&&&&&&&&&1&1&1&0 &1&1&1&1\\
\rule{0mm}{4mm}  
&&&&&1&&&&&&&1&1&1&1&0&1&1&1  \\
  &&&&&&1&&&&&&1&1&1&1&1&0&1&1 \\
  &&&&&&&1&&&&&1&1&1&1&1&1&0&1 \\
  &&&&&&&&1&&&&1&1&1&1&1&1&1&0 \\  
\end{array}
\right).
\label{GSpin32/Z2}
\eeqa
This generator matrix (\ref{GSpin32/Z2}) is obtained by 
multiplying the generator matrix 
\beqa
G_{\mbox{\scriptsize Pless\cite{Pless}}E_{16}}&=&
\left(
\begin{array}{cccc cccc cccc cccc}
  1&1&1&1& &&&&&&&&&&& \\
  &&1&1&1&1& &&&&&&&&&\\
  &&&&1&1&1&1& &&&&&&&\\
  &&&&&&1&1&1&1& &&&&&\\
\rule{0mm}{4mm}  
  &&&&&&&&1&1&1&1& &&&\\
  &&&&&&&&&&1&1&1&1& \\
  &&&& &&&& &&&& 1&1&1&1\\
  0&1&0&1&0&1&0&1&0&1&0&1&0&1&0&1
\end{array}
\right)
\label{GE16}
\eeqa
of the code $E_{16}$ given in the reference \cite{Pless}
by 
\beqa
\left(
\begin{array}{cccccccc}
 1 & 1 & 1 & 1 & 1 & 1 & 1 & 1 \\
  & 1 & 1 & 1 & 1 & 1 & 1 & 1 \\
  &  & 1 & 1 & 1 & 1 & 1 & 1 \\
  &  &  & 1 & 1 & 1 & 1 & 1 \\
  &  &  &  & 1 & 1 & 1 & 1 \\
  &  &  &  &  & 1 & 1 & 1 \\
  &  &  &  &  &  & 1 & 1 \\
  &  &  &  &  &  &  & 1 \\
\end{array}
\right)
\label{matrixleft}
\eeqa
from the left and 
\beqa
\left(
\begin{array}{cccccccccccccccc}
 1 &  &  &  &  &  &  &  &  &  &  &  &  &  &  &  \\
  &  &  &  &  &  &  &  & 1 &  &  &  &  &  &  &  \\
  & 1 &  &  &  &  &  &  &  &  &  &  &  &  &  &  \\
  &  &  &  &  &  &  &  &  & 1 &  &  &  &  &  &  \\
  &  & 1 &  &  &  &  &  &  &  &  &  &  &  &  &  \\
  &  &  &  &  &  &  &  &  &  & 1 &  &  &  &  &  \\
  &  &  & 1 &  &  &  &  &  &  &  &  &  &  &  &  \\
  &  &  &  &  &  &  &  &  &  &  & 1 &  &  &  &  \\
  &  &  &  & 1 &  &  &  &  &  &  &  &  &  &  &  \\
  &  &  &  &  &  &  &  &  &  &  &  & 1 &  &  &  \\
  &  &  &  &  & 1 &  &  &  &  &  &  &  &  &  &  \\
  &  &  &  &  &  &  &  &  &  &  &  &  & 1 &  &  \\
  &  &  &  &  &  & 1 &  &  &  &  &  &  &  &  &  \\
  &  &  &  &  &  &  &  &  &  &  &  &  &  & 1 &  \\
  &  &  &  &  &  &  &  &  &  &  &  &  &  &  & 1 \\
  &  &  &  &  &  &  & 1 &  &  &  &  &  &  &  &  \\
\end{array}
\right)
\label{matrixright}
\eeqa
from the right.
(\ref{matrixleft}) is a transformation of the lattice generator vectors, 
and (\ref{matrixright}) is a change of the lattice dimensions.
Here we show that the $D_{16}^+$ lattice can be constructed 
by Construction A from the codes generated by 
$G_{d_{16}^+}$ (\ref{GSpin32/Z2}).

In this case, the codeword $c\in\FF_2^{16}$ is obtained 
by left-multiplying $G_{d_{16}^+}$
by $x=(x_1\;x_2\;x_3\;x_4\;x_5\;x_6\;x_7\;x_8)\in\FF_2^8$ as
\beqa
x\,G_{d_{16}^+}=c=(c_1\;c_2\cdots c_{16}).
\eeqa
As we did in the extended Hamming code, 
let us consider separately the cases 
where $\sum_{i=1}^8 x_i$ is $\equiv0$ mod $2$ and $\equiv1$ mod $2$.

If $\sum_{i=1}^8 x_i\equiv0$ mod $2$, then codeword $c$ again
takes the form
\beqa
(x_1 \cdots x_8~~x_1 \cdots x_8),
\eeqa
whose symbols in the first half and the second half are the same.
Therefore, their contribution to the lattice theta function is
\beqa
\Theta_{x_1,1}^2\Theta_{x_2,1}^2\cdots\Theta_{x_8,1}^2.
\eeqa
According to (\ref{NSJacobi}), we have 
\beqa
&&\sum_{\sum_{i=1}^8 x_i\equiv0~\mbox{\scriptsize mod $2$}}
\Theta_{x_1,1}^2\Theta_{x_2,1}^2\cdots\Theta_{x_8,1}^2\n
&=&( \Theta_{0,1}^2)^8
+28( \Theta_{0,1}^2)^6( \Theta_{1,1}^2)^2
+70( \Theta_{0,1}^2)^4( \Theta_{1,1}^2)^4
+28( \Theta_{0,1}^2)^2( \Theta_{1,1}^2)^6
+( \Theta_{1,1}^2)^8
\n
&=&\frac12\left(
(( \Theta_{0,1}^2)^2+( \Theta_{1,1}^2)^2)^8+(( \Theta_{0,1}^2)^2-( \Theta_{1,1}^2)^2)^8
\right)\n
&=&\frac12\left(
(\vartheta_3^2)^8+(\vartheta_4^2)^8
\right)\n
&=&\frac12\left(
\vartheta_3^{16}+\vartheta_4^{16}
\right),
\label{Spin32/Z2NS}
\eeqa
obtaining the lattice theta when the 16 complex NS fermions are 
GSO-projected to the even fermion number sector,
in complete parallel to the case of the extended Hamming.

Similarly, for $\sum_{i=1}^8 x_i\equiv1$ mod $2$, the codeword $c$ is
\beqa
(x_1 \cdots x_8~~x_1+1 \cdots x_8+1),
\eeqa
which contribute to the lattice theta as 
\beqa
(\Theta_{x_1,1}\Theta_{x_1+1,1})
(\Theta_{x_2,1}\Theta_{x_2+1,1})
\cdots
(\Theta_{x_8,1}\Theta_{x_8+1,1})
&=&(\Theta_{0,1}\Theta_{1,1})^8.
\eeqa
Thus we have
\beqa
&&\sum_{\sum_{i=1}^8 x_i\equiv1~\mbox{\scriptsize mod $2$}}
(\Theta_{x_1,1}\Theta_{x_1+1,1})
(\Theta_{x_2,1}\Theta_{x_2+1,1})
\cdots
(\Theta_{x_8,1}\Theta_{x_8+1,1})\n
&=&
8(\Theta_{1,1}\Theta_{0,1})(\Theta_{0,1}\Theta_{1,1})^7
+56(\Theta_{1,1}\Theta_{0,1})^3(\Theta_{0,1}\Theta_{1,1})^5\n&&
+56(\Theta_{1,1}\Theta_{0,1})^5(\Theta_{0,1}\Theta_{1,1})^3
+8(\Theta_{1,1}\Theta_{0,1})^7(\Theta_{0,1}\Theta_{1,1})
\n
&=&\frac12
(\vartheta_2^{16}-\vartheta_1^{16}).
\label{Spin32/Z2R}
\eeqa
Setting $\vartheta_1=0$, this is 
exactly the theta of the 16 complex Ramond fermions 
of the Spin$(32)/\ZZ_2$ heterotic string.
Thus we have shown that 
the code generated by the generator matrix 
$G_{d_{16}^+}$ (\ref{GSpin32/Z2}) constructs 
the $D_{16}^+$ lattice by Construction A.

\vskip 5mm
\section*{Acknowledgments}
The authors thank S.~Yata for discussions.
The work of S.M. was supported by JSPS KAKENHI Grant Number JP23K03401,
and the work of T.O. was supported by JST SPRING, Grant Number JPMJSP2104.

\end{document}